\documentclass[twocolumn,showpacs,preprintnumbers,amsmath,amssymb]{revtex4}
\usepackage{graphicx}
\usepackage{bm}
\usepackage{amssymb} 

\begin{document}

\title{Finite-Size Effects in Surface-Enhanced Raman Scattering from
Molecules Adsorbed on Noble-Metal Nanoparticles}

\author{Vitaliy N. Pustovit and Tigran V. Shahbazyan}

\affiliation{Department of Physics and Computational Center for Molecular
  Structure and Interactions, Jackson State University, Jackson MS
  39217 USA}

\date{July 13, 2004} 

\begin{abstract}
We study the role of strong electron confinement in surface-enhanced
Raman scattering from molecules adsorbed on small noble-metal
particles. We describe a new source of Raman signal enhancement which
originates from different behavior of {\em sp}-band and {\em d}-band
electron densities near the particle boundary. In small
particles, a spillover of {\em sp}-electron wave-functions beyond the
classical radius gives rise to a thin layer with diminished population
of {\em d}-electrons. In this surface layer, the screening of 
{\em sp}-electrons by {\em d}-band electron background is reduced. We 
demonstrate that the interplay between finite-size and underscreening
effects results in an increase of the surface plasmon local field
acting on a molecule located in a close proximity to the particle
boundary. Our calculations, based on two-region model, show that the
additional enhancement of Raman signal gets stronger for smaller
nanoparticles due to a larger volume fraction of underscreened region. 
\end{abstract}


\maketitle

\section{INTRODUCTION}

\begin{figure}
\centering
\includegraphics[width=1.5in]{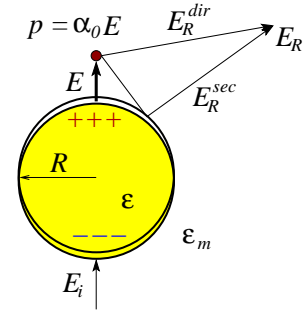}
\caption{\label{fig:sers-em}
Schematic representation of SERS from a molecule adsorbed on
a metal nanoparticle.
}
\end{figure}

A renewed interest in surface-enhanced Raman scattering
(SERS)\cite{fleischman-cpl74,vanduyne-jec77} stems from the discovery
of an enormous (up to $10^{15}$) enhancement of single-molecule Raman
signal in silver nanoparticle aggregates
\cite{nie-sci97,kneipp-prl97}. Although the relative importance of
various mechanisms of SERS is still an issue under active discussion,
the main source is attributed to the electromagnetic enhancement due
to the local field of surface plasmon (SP) excited in a nanoparticle
by the incident light
\cite{moskovits-jcp78,kerker-ao80,gersten-jcp80,schatz-review02} (see
Fig. \ref{fig:sers-em}).  Other possible enhancement mechanisms
involve dynamical charge transfer between a nanoparticle and a
molecule (chemical mechanism) and have been addressed, e.g., in Refs.
\cite{persson-cpl81,adrian-jcp82,moskovits-rmp85,otto-jpcm92}.  Recent
experimental
\cite{kneipp-cr99,brus-jacs99,brus-jpcb00,moskovits-tap02,rothberg-pnas04}
and theoretical
\cite{stockman-prb96,markel-prb96,xu-prb99,xu-prb00,stockman-prl03}
studies indicate that the anomalously strong Raman signal originates
from ``hot spots'' -- spatial regions where clusters of several
closely-spaced nanoparticles are concentrated in a small volume. The
high-intensity SERS then originates from the mutual enhancement of SP
local electric fields of several nanoparticles that determine the
dipole moment of a molecule trapped in a gap between metal surfaces.

Although in {\em single} nanoparticles the magnitude of SERS is
considerably smaller, with enhancement up to $10^6$ relative to the
Raman crossection of an isolated molecule, it varies substantially
with nanoparticles shape and size. In spheroidal particles, a strong
local field enhancement near the tip \cite{gersten-jcp80} leads to a
lightning rod effect recently observed in SERS from nanorods
\cite{elsayed-cpl02}. In gold nanorods \cite{feldmann-prl02} and
nanoshells \cite{halas-jcp99}, an additional local field enhancement
comes from a redshift of the SP energy away from the onset of optical
transitions between electronic {\em d}-band and {\em
sp}-band. However, in spherical Au particles, the proximity of SP and
interband transitions energies leads to a damping of SP by interband
electron excitations and, thus, to a reduction of the SP local
field. Such a damping is even stronger in Cu nanoparticles, where the
SP energy lies above the interband transitions onset
\cite{bigot-prl95}. In silver particles, however, the interband
transitions onset lies considerably higher in energy ($\sim 4.0$ eV)
than the SP ($\sim 3.0$ eV for nanoparticles in, e.g., glass matrix)
and have very litle detrimental effect on SERS.

Virtually all theoretical studies of the electromagnetic mechanism of
SERS were performed for relatively large nanoparticles with diameters
of several dozens nm or larger. For such sizes, the SP damping rate,
$\gamma$, in Ag particles comes mainly from the electron-phonon
scattering or electromagnetic retardation effects (for larger
particles). However, for particle sizes smaller than $\sim 10$ nm, the
finite-size effects become important. For small particles, the width
of the resonance peak in absorption is determined chiefly by the SP
damping due to excitation of high-energy {\em intraband}
single-particles transitions accompanied by transfer of momentum to
nanoparticle boundary \cite{kubo-jpsj66}. This effect is usually
incorporated via size-dependent correction,
$\gamma=\gamma_0+\gamma_s$, in the Drude dielectric function for 
{\em sp}-electrons,
\begin{equation}
\label{s-drude}
\epsilon_s(\omega)=1-\omega_p^2/\omega (\omega+i\gamma),
\end{equation}
where $\omega_p$ is the bulk plasmon frequency ($\omega_p\simeq 9.0$
eV for Ag), and $\gamma_s$ is determined by the electron level spacing
near the Fermi level,
\begin{equation}
\label{width}
\gamma_s\sim v_F/R,
\end{equation}
with a numerical coefficient of the order of unity (we set
$\hbar=1$). For small particles, $\gamma_s$ dominates over the
phonon-induced damping, $\gamma_0$, resulting in a strong dependence
of SERS on nanoparticle size. Indeed, the amplitude of the SP local
field {\em at resonance} is $E\propto \gamma^{-1}$, so that the
electromagnetic enhancement factor\cite{kerker-ao80} depends on the
radius as $A(\omega)=|E/E_i|^4\propto R^4$ (here $E_i$ is the incident
field). Thus, in small particles, finite size effects can reduce SERS
by several orders of magnitude.

In this paper, we demonstrate that for even smaller nanometer-sized
particles, another quantum-size effect, with the opposite trend
towards {\em increasing} SERS, becomes important. The underlying
mechanism is related to the difference in the density profiles of 
{\em sp}-band and {\em d}-band electrons near the nanoparticle
boundary. Specifically, the {\em localized} {\em d}-electrons are
mainly confined within nanoparticle classical volume while the
wave-functions of {\em delocalized} {\em sp}-electrons extend outside
of it. This spillover leads to a larger {\em effective} radius for
{\em sp}-electrons \cite{persson-prb85} and, thus, to the existence of
a surface layer with diminished {\em d}-electron population
\cite{liebsch-prb93,kresin-prb95,liebsch-prb95}. As a result, in the
surface layer, the screening of {\em sp}-electrons by {\em d}-band
electron background is reduced, leading to a blueshift of SP
absorption peak in small Ag nanoparticles. More recently, the effect
of underscreening of Coulomb interactions between {\em sp}-electrons
has been observed as an enhancement of electron-electron scattering
rate measured in ultrafast pump-probe spectroscopy
\cite{voisin-prl00,liebsch-prb01} and photoemission
\cite{freund-prl00} experiments.

Specifically, we address the role underscreening plays in the SERS
from a molecule located in a close proximity to the surface of an Ag
nanoparticle. We find a substantial increase of the SP local field
{\em outside} of nanoparticle and, hence, an additional enhancement of
the Raman signal. Furthermore, we investigate the size-dependence of
SERS to find that the enhancement factor deviates considerably from
the $R^4$ behavior when the screening effects are included. In
particular, the relative enhancement is stronger for smaller
nanoparticles due to a larger volume fraction of the underscreened
region.

The paper is organized as follows. In Section \ref{sec-model} we
describe the two-region model which we adopt to incorporate the
surface layer effect. In Section \ref{sec-raman} we calculate local
fields and Raman enhancement factor. Discussion of our numerical
results is presented in Section \ref{sec-disc}. Section \ref{sec-conc}
concludes the paper.

\section{MODEL}
\label{sec-model}

We consider SERS from a molecule adsorbed on a Ag nanoparticle in a
medium with dielectric constant $\epsilon_m$. For nanoparticle
diameters exceeding $\approx 1.5$ nm, the bulk electronic structure is
essentially preserved while the quantum-mechanical corrections due to
discreteness of the electron energy spectrum can be included via
$\gamma_s$ in the {\em sp}-electron dielectric function
Eq. (\ref{s-drude}). To incorporate the surface layer effect, we adopt
the two-region model where {\em sp}-band and {\em d}-band electrons
are confined within spherical volumes with different radii $R_d$ and
$R$, respectively \cite{lushnikov-zf77,liebsch-prb93,kresin-prb95}.
Note that for {\em d}-band electrons, the semiclassical approach
\cite{lushnikov-zf77} remains valid even for smaller nanometer-sized
particles \cite{kresin-prb95}, while for {\em sp}-electrons, the
density $n_s(r)$ is a smooth function near the boundary while
exhibiting Friedel oscillations inside the nanoparticles
\cite{ekardt-prb85}. A fully quantum-mechanical theory of SERS in
nanometer-sized particles will be published elsewhere
\cite{pustovit-prl04}. For nanoparticles under consideration, however,
the deviations of {\em sp}-electron density from classical shape do
not affect the situation qualitatively \cite{liebsch-prb93} and
$n_s(r)$ can be approximated by a step-function with a sharp boundary
at the effective radius $R$.

The frequency-dependent potential is determined from the Poisson equation,
\begin{equation}
\label{poisson}
\Phi(\omega, {\bf r})=\phi_0({\bf r}) + \int d^3r'
\frac{\delta N(\omega,{\bf r}')}{|{\bf r}-{\bf r}'|},
\end{equation}
where $\phi_0({\bf r})=-e{\bf E}_i\cdot {\bf r}$ is the potential of
incident light with electric field amplitude ${\bf E}_i=E_i
{\bf z}$ along the $z$-axis, and $\delta N(\omega,{\bf r})$ is the
induced charge density (hereafter we suppress the frequency
dependence). The latter has four contributions,
\begin{equation}
\label{ind-density}
\delta N({\bf r})=\delta N_s({\bf r})
+\delta N_d({\bf r})+\delta N_m({\bf r})+ \delta N_0({\bf r}),
\end{equation}
originating from the valence {\em sp}-electrons, 
the core {\em d}-electrons, 
the dielectric medium, 
and the molecule, 
respectively.
The density profile of delocalized {\em sp}-electrons is not fully
inbedded in the background of localized {\em d}-electrons but extends
over it by $\Delta = R-R_d$ which, within our model, is the surface
layer thickness \cite{lushnikov-zf77,liebsch-prb93,kresin-prb95}. The
induced density is expressed via electric polarization vector as
%
\begin{equation}
\label{el-pol}
\delta N({\bf r})=-\nabla \cdot {\bf P}({\bf r})
=-\nabla \cdot ({\bf P}_d+{\bf P}_s+{\bf P}_m+{\bf P}_0),
\end{equation}
with each contribution related back to the potential as
\begin{eqnarray}
\label{el-pols} 
{\bf P}_d({\bf r}) 
&=&
-\frac{\epsilon_d-1}{4\pi} \,
\theta(R_d-r)\nabla \Phi({\bf r}),
\nonumber\\
{\bf P}_s({\bf r}) 
&=&
-\frac{\epsilon_s-1}{4\pi} \,
\theta(R-r)\nabla \Phi({\bf r}),
\nonumber\\
{\bf P}_m({\bf r}) 
&=&
-\frac{\epsilon_m-1}{4\pi} \,
\theta(r-R)\nabla \Phi({\bf r}),
\nonumber\\
 {\bf P}_0({\bf r}) 
&=&
- \,\delta({\bf r}-{\bf r}_0) \, \alpha_0 \nabla \Phi({\bf r}_0),
\end{eqnarray}
where step functions $\theta(x)$ enforce the corresponding boundary
conditions. The molecule is represented by a point dipole with
polarizability $\alpha_0$ located at ${\bf r}_0$ (we chose
nanoparticle center as the origin). In the following, averaging over
the orientations of molecular dipole is implied so the polarizability
tensor $\alpha_0$ is assumed to be isotropic.

After substituting above expressions into r.h.s. of Eq.\ (\ref{poisson}) and
integrating by parts, we obtain a self-consistent equation for the potential 
$\Phi({\bf r})$,
\begin{eqnarray}
\label{poisson1}
\epsilon(r)\Phi({\bf r}) 
&=&
\phi_0({\bf r}) 
-\nabla_0\frac{1}{|{\bf r}-{\bf r}_0|} 
\cdot \alpha_0\nabla_0\Phi({\bf r}_0)
\nonumber\\
&+&
\int d^3r' \nabla'\frac{1}{|{\bf r}-{\bf r}'|}\cdot 
\nabla'[\chi_d\theta(R_d-r')
\nonumber\\
&+&
\chi_s\theta(R-r')+\chi_m\theta(r'-R)]
\Phi({\bf r}'),
\end{eqnarray}
where $\epsilon(r)= \epsilon_d+\epsilon_s-1$, $\epsilon_s$, and
$\epsilon _m$ in the intervals $r<R_d$, $R_d<r<R$, and $r>R$,
respectively, and $\chi_i=(\epsilon_i-1)/4\pi$ ($i=s,d,m$) are the
corresponding susceptibilities. For simplicity, we assume that the
molecule is located on the $z$-axis (along incident field direction).
After expanding $\Phi({\bf r})$ and $|{\bf r}-{\bf r}'|^{-1}$ in terms
of spherical harmonics and  retaining 
only the dipole terms, we obtain for the radial component
\begin{eqnarray}
\label{poisson2}
\epsilon(r)\Phi(r) &=& 
\phi_0(r) 
- \frac{\epsilon_d-1}{3}\, \beta \biggl(\frac{r}{R_d} \biggr)
\,\Phi(R_d)
\nonumber\\
&+& \frac{\epsilon_m-\epsilon_s}{3}\, \beta \biggl(\frac{r}{R} \biggr)
\,\Phi(R)
\nonumber\\
&-& \frac{4\pi \alpha_0}{3r_0^2}\, \beta \biggl(\frac{r}{r_0} \biggr)
\, \frac{\partial \Phi(r_0)}{\partial r_0},
\end{eqnarray}
where 
\begin{equation}
\label{beta}
\beta(x) = x^{-2}\, \theta (x-1) -2x \theta(1-x).
\end{equation}

The second and third terms in rhs of Eq. (\ref{poisson2}) describe
light scattering from the boundaries at $r=R_d$ and $r=R$,
respectively, that separate regions with different dielectric
functions, while the last term represents the potential of the
molecular dipole.  The boundary values of $\Phi$ are found by setting
$r=R_d, R, r_0$ in Eq. (\ref{poisson2}), which leads to a closed-form
expression for the self-consistent potential in the presence of
molecule, nanoparticle, and dielectric medium.

\section{CALCULATION OF RAMAN SIGNAL}
\label{sec-raman}

The dipole moment of a radiating molecular dipole is determined by the
local electric field, {\bf E}, at the molecule location: ${\bf
p}=\alpha_0 {\bf E}$. The Raman field consists of a direct field of
this dipole and a secondary field scattered by the nanoparticle. In
order to extract the Raman signal, we present the self-consistent
potential in the form $\Phi=\phi+\phi^R$, where $\phi$ is the local
potential in the absence of molecule and $\phi^R$, calculated in the
first order in $\alpha_0$, determines the Raman signal
\cite{schatz-review02}.

Keeping only zero-order terms in Eq. (\ref{poisson2}), we find for the
nanoparticle local potential
\begin{eqnarray}
\label{phi-class}
\phi=\varphi_0+\delta \varphi,
~~~
\varphi_0=\phi_0/\epsilon(r)=-E_ir/\epsilon(r),
\nonumber\\
\delta \varphi(r) = \frac{1}{\epsilon(r)} 
\Bigl[ - \beta(r/R_d) \phi_0(R_d)\, 
\frac{\lambda_d (1-2\lambda_m)}{1-2a^3\lambda_d\lambda_m}
\nonumber\\
 + \beta(r/R) \phi_0(R)\,
\frac{\lambda_m (1-a^3\lambda_d)}{1-2a^3\lambda_d\lambda_m}\Bigr],
\end{eqnarray}
where $a=R_d/R$ is the ``aspect ratio'', and the parameters $\lambda_d$ and
$\lambda_m$ are given by
\begin{eqnarray}
\label{lambda-class}
\lambda_d=\frac{\epsilon_d-1}{\epsilon_d+3\epsilon_s-1}, ~~~
\lambda_m=\frac{\epsilon_m-\epsilon_s}{2\epsilon_m+\epsilon_s}.
\end{eqnarray}
The spatial dependence of induced potential,
$\delta\varphi$, is determined by $\beta(x)$ of
Eq. (\ref{beta}). Inside the particle, the 
potential linearly increases for $r<R_d$, while it exhibits a more
complicated behavior in the surface layer, $R_d<r<R$. Outside the
nanoparticle, $\delta\varphi$ falls off quadratically, 
\begin{equation}
\label{delta-phi-far}
\delta \varphi(r) = \frac{E_i \alpha}{\epsilon_m r^2},
~~~ r>R,
\end{equation}
where $\alpha(\omega)$ is the particle polarizability 
\begin{eqnarray}
\label{alpha-class}
\alpha(\omega)=
R^3\, \frac{a^3\lambda_d (1-\lambda_m)-\lambda_m}{1-2a^3\lambda_d\lambda_m}.
\end{eqnarray}
In the absence of the surface layer, $R_d=R$, we recover the usual
expression 
%
\begin{equation}
\label{alpha-class-0}
\alpha^0=
R^3\,
\frac{\epsilon_s+\epsilon_d-1-\epsilon_m}
{\epsilon_s+\epsilon_d-1+2\epsilon_m},
\end{equation}
with the resonance at
$\omega_M=\frac{\omega_p}{\sqrt{\epsilon_d+2\epsilon_m}}$. 
From Eqs.\ (\ref{phi-class}) and (\ref{delta-phi-far}), 
we then obtain the local electric field at the molecule 
location as
\begin{equation}
\label{e-field-class}
E(r_0)=-\frac{\partial \varphi}{\partial r_0} 
= \frac{E_i}{\epsilon_m} \, (1+2g),
~~
g(\omega)=\frac{\alpha(\omega)}{r_0^3}.
\end{equation}
To calculate the Raman signal, we substitute the local field into the
last term of Eq. (\ref{poisson2}) and, in the first order in
$\alpha_0$, obtain the following equation for $\phi^R$,
\begin{eqnarray}
\label{poisson-raman}
\epsilon(r)\phi^R(r) 
&=& \phi_0^R(r) 
- \frac{\epsilon_d-1}{3}\, \beta(r/R_d)\phi^R(R_d)
\nonumber\\
&+&
\frac{\epsilon_m-1}{3}\, \beta(r/R) \phi^R(R),
\end{eqnarray}
where the potential
\begin{eqnarray}
\label{phi-r-0}
\phi_0^R(r)= 
\frac{4\pi\alpha'_0}{3r_0^2}\,  E(r_0)\,
\Biggl[\frac{r_0^2}{r^2}\,\theta(r-r_0)
-\frac{2r}{r_0}\theta(r_0-r)\Biggr],
\end{eqnarray}
describes the direct field of a radiating molecular dipole
($\alpha'_0$ is the derivative of molecule polarizabilty with respect
to normal coordinate that determines the Stokes shift $\omega_S$
\cite{long-book}), while the second and third terms describe secondary
fields due to scattering from {\em d}-band and {\em sp}-band electron
distributions boundaries, respectively. The latter can be found by
matching $\phi^R$ at $r=R_d$ and $r=R$. We now notice that, at these
values, $\phi_0^R(r)$ is linear in $r$, so we can write
$\phi^R=\varphi_0^R+\delta \varphi_0^R$ with 
\begin{eqnarray}
\varphi_0^R=\phi_0^R/\epsilon(r),
~~
\label{delta-phi-r}
\delta \phi^R(r)= 
\frac{8\pi\alpha'_0}{3r_0^3}\,  E(r_0)\, \delta \varphi(r),
\end{eqnarray}
where $\delta \varphi(r)$ is given by Eq. (\ref{phi-class}) but with
$\omega_S$ instead of $\omega$. We then finally
obtain for the Raman field ($r>r_0)$
\begin{equation}
\label{phi-raman-far}
\phi^R(r) = - \frac{4\pi \alpha'_0 E_i}{3\epsilon_m r^2}
\Bigl[1+2g(\omega)\Bigr]\Bigl[1+2g(\omega_s)\Bigr],
\end{equation}
and, hence, for the enhancement factor
\begin{equation}
\label{raman-factor}
A(\omega,\omega_s)=
\Bigl|1+2g(\omega)+2g(\omega_s)+ 4g(\omega)g(\omega_s)\Bigr|^2.
\end{equation}
The above expressions generalize the well-known classical result
\cite{kerker-ao80,gersten-jcp80,schatz-review02} to the case of a
small noble-metal particle with different profiles of {\em d}-band and
{\em sp}-band densities. While SERS retains the usual dependence on
nanoparticle polarizability, the latter is modified in the presence of
a surface layer [see Eq. (\ref{alpha-class})]. Note that
Eq. (\ref{phi-raman-far}) remains unchanged even for non-classical
electron distributions provided that electronic wave-functions in a
nanoparticle do not overlap with molecular orbitals
\cite{pustovit-prl04}.


\section{NUMERICAL RESULTS}
\label{sec-disc}

Below we present the results of numerical calculations for Ag
nanoparticles with diameters ranging from 2 to 6 nm in a medium with
dielectric constant $\epsilon_m=2.0$. The SP resonance is positioned
at $\omega_M\simeq 3.0$ eV, far away from the interband transitions
onset in Ag at 4.0 eV, so in the frequency range of interest the real
part of interband dielectric function is nearly a constant,
$\epsilon_d\simeq 5.2$. In this size range, the SP damping is
dominated by size-dependent contribution to $\gamma$, with the
numerical coefficient in $\gamma_s$ [see Eq. (\ref{width})] adjusted
to fit the experimental absorption data \cite{voisin-prl00}.

\begin{figure}[tb]
 \centering
 \includegraphics[width=3.5in]{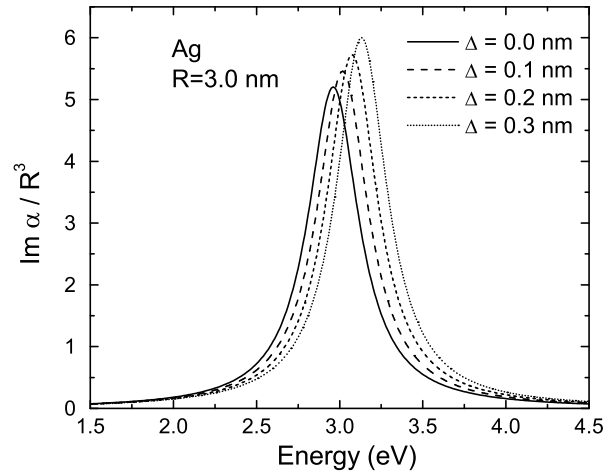}
  \caption{\label{fig:abs-class}
Absorption spectra of Ag nanoparticle calculated for $R=3.0$ nm and
several values of surface layer thickness $\Delta$.
}  
 \end{figure}

In Fig. \ref{fig:abs-class}, we show calculated absorption spectra for
various surface layer thicknesses $\Delta$. For finite thicknesses,
the SP energy experiences a blueshift whose magnitude increases with
$\Delta$, in agreement with previous calculations of absorption in
small silver particles \cite{liebsch-prb93,kresin-prb95}. This
blueshift originates from a reduction of the {\em effective} (averaged
over the volume) interband dielectric function in the nanoparticle
that determines the SP energy. At the same time, the peak amplitude
increases with $\Delta$ while the resonance width is unchanged. The
absolute value of polarizability determines, in turn, the magnitude of
the local field outside the nanoparticle [see
Eq. (\ref{e-field-class})].
%
%

\begin{figure}[b]
 \centering
 \includegraphics[width=3.5in]{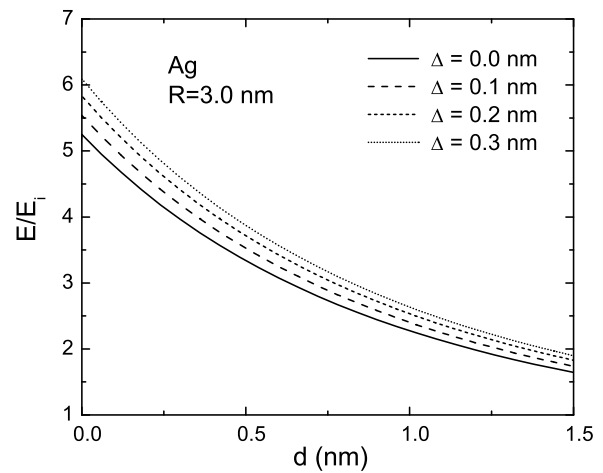}
 \caption{\label{fig:e-field-class}
Local field at surface plasmon energy as a function of distance from
particle boundary calculated for $R=3.0$ nm and several values of
surface layer thickness $\Delta$. 
} 
 \end{figure}

In Fig. \ref{fig:e-field-class} we plot the local electric field at SP
resonance frequency as a function of molecule distance from the metal
surface, $d=r-r_0$. Outside the nanoparticle, the local field exhibits
the usual $r^{-3}$ decay relative to constant incident field
background.  At the same time, the field magnitude is larger for
finite $\Delta$, reaching $\simeq 20\%$ enhancement near the boundary
for 3.0 {\AA} thick surface layer. Enhancement is strongest when the
molecule is located in a close proximity to metal surface, near the
{\em underscreened} region with small population of {\em d}-electrons.
%
%

\begin{figure}[tb]
 \centering \includegraphics[width=3.5in]{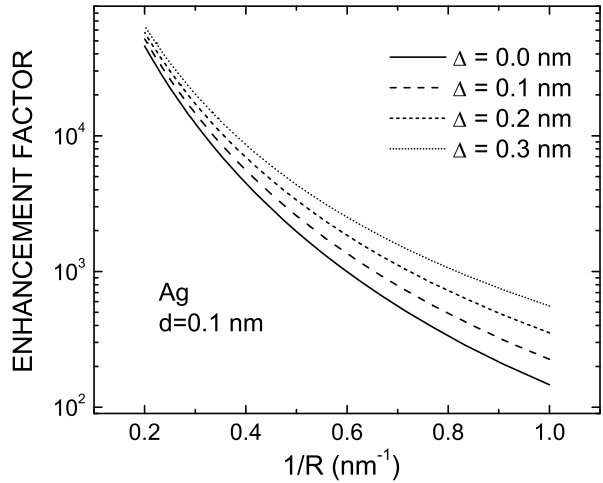}
 \caption{\label{fig:enhance-class}
   Size-dependence of Raman enhancement factor calculated for
   particle-molecule distance $d=0.1$ nm and several values of surface
   layer thickness $\Delta$.  }
 \end{figure}

We now turn to the size-dependence of SERS. In
Fig. \ref{fig:enhance-class}, we plot the enhancement factor at SP
energy as a function of particle radius for a small molecule-particle
separation $d=0.1$ nm. We consider here non-resonant scattering and
assume, as usual, that molecular vibrational energies lie within SP
resonance width (the latter being large for nanometer-sized
particles). In this case, the main contribution to SERS comes from the
last term of Eq. (\ref{raman-factor}) corresponding to the secondary
scattered field of radiating molecular dipole. As it can be seen in
Fig. \ref{fig:enhance-class}, the general tendency is a decrease of
the Raman signal for small nanoparticles due to a strong SP damping by
single-particle excitations. For $\Delta=0$, the size-dependence of
enhancement factor is $A \propto R^4$. However, that dependence
changes considerably when the effect of surface layer is included in
the calculations. For finite $\Delta$, the decrease of $A$ is
considerably slower, with the signal strength in small nanoparticles
about 400\% larger than that for $\Delta=0$. It should be emphasized
that the role of the surface layer becomes more pronounced when
particle size is decreased. Indeed, the {\em relative} enhancement of
SERS is larger for smaller particles [see
Fig. \ref{fig:enhance-class}] due to a larger volume fraction of the
underscreened region.

%
%

\section{CONCLUSIONS}
\label{sec-conc}

We investigated the role of quantum-size effects in the
electromagnetic mechanism of surface-enhanced Raman scattering in
noble-metal nanoparticles. We identified a new source of Raman signal
enhancement in small particles which originates from different density
profiles of {\em sp}-band and {\em d}-band electrons near the
boundary. The existence of an underscreened surface layer with low
population of {\em d}-electrons gives rise to a stronger, compared to
classical calculations, local field of surface plasmon collective
excitation that determines the magnitude of Raman signal from a
molecule near the surface. Although the dominant finite-size effect is
still a reduction of SERS due to surface plasmon damping, the decrease
of the Raman signal is considerably slower when the surface layer
effect is taken into account.

We calculated the size-dependence of SERS in the framework of two
region model with different (but classical) distributions of 
{\em sp}-band and {\em d}-band electron densities and semiclassical
treatment of electron surface scattering, and found that the
additional enhancement becomes stronger as the particle radius
decreases. Although for particles size smaller than 2-3 nm, the
semiclassical model is no longer valid, the physical mechanism of the
enhancement persists for even smaller sizes. In fact, the
semiclassical model understimates the fraction of underscreened region
as the particle size decreases. On the other hand, in the absence of a
sharp boundary the local fields are weaker. The outcome of this
competition depends on the precise shape of the electron density as
well as on the surrounding dielectric. A fully quantum-mechanical
calculations of SERS will be presented in a subsequent publication
\cite{pustovit-prl04}.

This work was supported by NSF under grants DMR-0305557 and
NUE-0407108, by NIH under grant 5 SO6 GM008047-31, and by ARL under
grant DAAD19-01-2-0014. TVS thanks Max-Planck-Institut f\"{u}r
Physik Komplexer Systeme for the hospitality.


\end{document}